 \documentstyle[12pt]{article}
 \oddsidemargin--1mm
\begin{document}
 
\begin{center}
 {\LARGE The Inverse Variational Problem for Autoparallels}

 \vskip 0.5cm
Christian MAULBETSCH and
 Sergei V. SHABANOV \footnote{On leave from Laboratory 
 of Theoretical
 Physics, JINR, Dubna, Russia.}

 \vskip 0.5cm
 {\em Institute for Theoretical Physics, FU-Berlin, Arnimallee 14,
 Berlin,\\ D-14195, Germany }
 \end{center}

\newcommand{\pll}{\partial}
\renewcommand{\theequation}{\thesubsubsection.\arabic{equation}}
\renewcommand{\thesubsubsection}{\arabic{subsubsection}}

\newcommand{\be}{\begin{equation}}
\newcommand{\ee}{\end{equation}}
\newcommand{\bea}{\begin{eqnarray}}
\newcommand{\eea}{\end{eqnarray}}
\newcommand{\og}{\overline{\Gamma}}
\newcommand{\ts}{\tilde{S}}
\newcommand{\e}{\epsilon}
\newcommand{\la}{\lambda}
\newcommand{\p}{\partial}
\newcommand{\s}{\sigma}

\begin{abstract}
 We study the problem of the existence of  a local 
quantum scalar field theory in a general affine metric
space that in the semiclassical approximation would
lead to the autoparallel motion of wave packets, thus
providing a deviation of the spinless particle trajectory
from the geodesics in the presence of torsion.
The problem is shown to be equivalent to the inverse
problem of the calculus of variations for the autoparallel
motion with additional conditions that the action (if it 
exists) has to be invariant under time reparametrizations
and general coordinate transformations, 
while depending analytically
on the torsion tensor.
The problem is proved to have no  solution for a generic torsion in
four-dimensional spacetime.
A solution exists only if the contracted torsion tensor is
a gradient of a scalar field.
The corresponding field theory describes coupling of matter to the dilaton
field.
\end{abstract}

\subsubsection{Introduction and motivations}
\setcounter{equation}0

In Riemann-Cartan spaces, a connection ${\Gamma_{\mu\nu}}^{\s}$
compatible with the metric $g_{\mu\nu}$ (meaning that 
$D_{\mu} g_{\nu\s} = 0$, with $D_{\mu}$ being 
the covariant derivative) 
may have nonvanishing antisymmetric components 
${S_{\mu\nu}}^{\s}= \frac{1}{2} ({\Gamma_{\mu\nu}}^{\s} -
{\Gamma_{\nu\mu}}^{\s})$ which are the torsion tensor components
in a coordinate basis.
A general affine connection compatible with the metric can always
be represented in the form \cite{schouten}
${\Gamma_{\mu\nu}}^{\s} = {\og_{\mu\nu}}^{\s} +
g^{\s\alpha}( S_{\mu\nu\alpha}-S_{\nu\alpha\mu}+S_{\alpha\mu\nu})$ ,
where ${\og_{\mu\nu}}^{\s}$ are the Christoffel symbols associated
with the metric $g_{\mu\nu}$.
As was first pointed out by Cartan, the existence of connections that
are compatible with the metric and do not coincide with the
natural Riemannian connection ${\og_{\mu\nu}}^{\s}$ may lead
to more general theories of gravity than Einstein's 
general relativity (see, e.g., for a review \cite{hehl95} 
and references therein).
Consequently, the actual motion of a spinless point particle
may, in principle, deviate from the usual geodesic motion due to an
interaction with torsion.

The torsion force can not be arbitrary and its possible form should
be obtained from some physical principles.
It is natural to assume the actual motion of a particle 
to enjoy  general coordinate covariance.
A trajectory of the motion is determined by its tangent vector
(or velocity). So to specify the corresponding equations of motion,
one has to define the variation of the velocity along the trajectory.
In a space with a general affine connection there exist two independent
variation operators that involve a displacement and produce tensors
out of tensors (i.e., variations covariant under general
coordinate transformations): the Lie derivative and the covariant 
derivative \cite{schouten}, p.335. 
A physically acceptable variation should contain the displacement
$d_{u} u^{\mu} = \dot{u}^{\mu}$ of the velocity along itself 
(acceleration). The Lie derivative does not provide 
us with such a displacement.
Therefore the only possibility is
\be
\label{auto}
D_{u} u^{\mu} = u^{\nu} D_{\nu} u^{\mu} =
 \dot{u}^{\mu} + {\Gamma_{\nu\s}}^{\mu} u^{\nu} u^{\s} =
F^{\mu} ( S,g,u)\, ,
\ee
where $F^{\mu}$ is a vector force.
Next we require that the motion becomes geodesic when the torsion vanishes,
that is, the vector $u^{\mu}$ is transported 
parallel along itself with respect to a natural Riemannian
connection ${\og_{\mu\nu}}^{\s}$
\be
\label{geo}
\overline{D}_{u} u^{\mu} = 
\dot{u}^{\mu} + {\og_{\nu\s}}^{\mu} u^{\nu} u^{\s} = 0 \, .
\ee
This implies the condition $F^{\mu} (S=0,g,u) = 0$.
The simplest possibility proposed first by Ponomarev \cite{ponomarev}
is to set $F^{\mu}=0$.
The corresponding curve is called the autoparallel.
Its characteristic geometrical property is similar to that of geodesics.
The tangent vector is transported parallel along itself with
respect to a full affine connection.
But it does not share another property of geodesics such as being the
shortest line between two points of the manifold.
 
As follows from the comparison of Eqs. (\ref{geo}) and (\ref{auto})
with $F^{\mu}=0$, the deviation of the autoparallel from the geodesic
is caused by the torsion force $2 \, S_{\mu\nu\s} u^{\nu} u^{\s}$.
The choice between the geodesic and the autoparallel motion can either
be decided experimentally or on theoretical grounds following from
the compatibility of the postulate $F^{\mu}=0$ in (\ref{auto})
with other fundamental principles of physics.
In \cite{hehl71} it is argued that the energy-momentum conservation law 
of a spinless point particle leads to geodesics rather than to
autoparallels. The conclusion is based on the earlier work by 
Papapetrou \cite{papa} that prescribes a specific relation between 
the canonical momentum and the velocity of the particle.
In general, the energy-momentum tensor is defined as the variational 
derivative of the particle Lagrangian with respect to the metric tensor.
Its conservation law specifies the particle equations of motion
that are the usual Euler-Lagrange equations.
Hence, if equation (\ref{auto}) admits the Euler-Lagrange form, then 
the energy momentum conservation law may lead to the autoparallels
as is shown in Appendix A with an explicit example.

Based on a physical analogy between spaces with torsion and 
crystalls with topological defects \cite{kondo},
the attention has been brought again to
the autoparallel motion in \cite{kleinertbook}, Sec. 10, where it was also
quantized by the path integral method. 
The approach gives a consistent quantum theory 
only for a special (``gradient'')
torsion \cite{kleinertbook}, Sec. 11. For a generic torsion it has
lead to difficulties with the probabilistic interpretation
of the corresponding quantum mechanics
and with the correspondence principle \cite{arik}.

The problem of coupling between matter and the spacetime geometry
is undoubtedly of great importance.
So far only the principle of minimal gauge coupling has been explored
\cite{kibble,hehl95}, except, maybe, for the conformal coupling 
\cite{birrell}.
The aim of the present work is to approach the problem from a different
and more general point of view. All models of the fundamental interactions
are described by quantum field theory.
Thus,  if the autoparallels indeed describe the motion of a spinless
point particle in a general Riemann-Cartan space, then they must follow
from a local quantum scalar field theory in the semiclassical (eikonal)
approximation. A conventional way to construct a quantum field theory
that satisfies the correspondence principle is first to quantize 
the relativistic particle motion, thus obtaining relativistic 
quantum mechanics, and then to apply the so called second
quantization procedure \cite{bjorken}.

Consider, for example, the geodesic motion (\ref{geo}).
It follows from a least action principle for the action
\be
\label{actgeo}
S_{g} = \int \! L_{g} d t = - m \int \! \sqrt{g_{\mu\nu}v^{\mu}v^{\nu}}d t =
-m \int \! d s  \, ,   
\ee
where $v^{\mu}= d q^{\mu} /d t$.
In (\ref{auto}) it has been set $ \dot{u}^{\mu}=d u^{\mu}/d s$
and $u^{\mu}= d q^{\mu}/d s$.
To quantize the system, one goes over to the canonical
Hamiltonian formalism by means of the Legendre transformation for $v^{\mu}$.
Defining the canonical momentum $p_{\mu}= \p L_{g}/ \p v^{\mu}$
we find that the canonical Hamiltonian $H=p_{\mu} v^{\mu} - L_{g} = 0$
vanishes identically.
This happens due to the local time reparametrization symmetry
of the action (\ref{actgeo}).
It is not hard to be convinced that the Hessian 
$H_{\mu\nu}= \p^2 L_{g}/(\p v^{\mu} \p v^{\nu})$ is degenerate
(in particular, $H_{\mu\nu} v^{\nu} = 0$)
and, therefore, the system has a constraint.
It has the well known form
$\Pi = p^{2} - m^{2} =0$.
According to Dirac \cite{diraclec}, after promoting $p_{\mu}$ and $q^{\mu}$
to self-adjoint operators satisfying the Heisenberg algebra,
the constraint $\hat{\Pi}$ has to annihilate physical states
\be
\label{conqua}
\hat{\Pi} \psi = ( \hat{p}^{2}- m^{2} ) \psi = 0 \, ,
\ee
where $-\hat{p}^{2}$ is the Laplace-Beltrami operator ($\hbar=1$).
In doing so, we have obtained a relativistic quantum mechanics that 
leads to the geodesic motion of the wave packets in the eikonal
approximation.
Note that the canonical Hamiltonian vanishes identically, hence,
the Schr\"odiger evolution 
$i \p_{t} \psi = \hat{H} \psi \equiv 0$ is trivial.
So, the constraint (\ref{conqua})
entirely specifies the evolution of relativistic quantum particle states.
This latter property allows one to construct a corresponding
quantum field theory. 
If all solutions of (\ref{conqua}) are labelled by 
a set of parameters $k$,
then a Heisenberg quantum field operator that carries quanta (particles)
with quantum numbers $k$ and wave functions $\psi_{k}(q)$
reads $\hat{\phi}= \Sigma_{k} \psi_{k}(q) \hat{a}_{k} +  h.c.$ 
where $\hat{a}_{k}$ and $\hat{a}^{\dagger}_{k}$ are destruction and creation
operators of these quanta. 
The corresponding action of such a field theory 
in $n$ dimensions is \cite{birrell}
\be
\label{qftact} 
S= \int d^{n} q  \sqrt{g} \; \phi \, \hat{\Pi} \, \phi 
= \int d^{n} q  \sqrt{g} \left( g^{\mu\nu} \p_{\mu} \phi
 \p_{\nu} \phi
- m^{2} \phi^{2} \right) \, .
\ee
Thus the constraint occuring through the time
reparametrization (gauge) symmetry specifies the sought-for quantum field
theory  obeying the  correspondence principle.

The same strategy could be applied to build a relativistic quantum theory for
the autoparallel motion.
That is, we need a Lagrangian for the equation (\ref{auto}).
It has to fulfill some additional physical conditions:
{\em (i)} to be time reparametrization invariant,
{\em (ii)} to be invariant under general coordinate transformations 
(i.e., to be a scalar) and
{\em (iii)} to turn into (\ref{actgeo}) as the torsion approaches zero
(analyticity in torsion).
We remark that the autoparallel equation (\ref{auto}) 
with $F_{\mu} \equiv 0$ 
exhibits the time reparametrization symmetry therefore it is natural 
to expect the Lagrangian to fulfill the condition {\em (i)}.
Yet, as has been pointed out, the constraint occuring through
this gauge symmetry entirely determines the evolution
of a {\em relativistic} quantum particle interacting with the
spacetime geometry.
The second condition is the standard one:
Physics can not depend on the choice of a coordinate system.
The third one is natural since we expect a small deviation from
the geodesic motion in the limit of small torsion.
\smallskip
So, we have reduced our problem to the well-known and, in fact,
long-standing problem of mathematical physics:
given a set of equations of motion, find out whether they admit 
the Euler-Lagrange form.
This is the inverse problem of the calculus of variations.
Necessary and sufficent conditions for the solution to exist have
been first formulated by Helmholtz \cite{helmholtz}.

\subsubsection{The Helmholtz conditions for the autoparallel motion}
\setcounter{equation}0

Let the equations of motion be a system of differential equations 
of second order
\be
G_{\mu} (\dot{v},v,q) = 
H_{\mu\nu} (v,q) \dot{v}^{\nu} + B_{\mu}(v,q) = 0.
\ee
The question arises:
Does there exist a Lagrangian whose Lagrange derivative $[L]_{\mu}$
coincides with the equation of motion? That is,
\be
\label{lg}
G_{\mu} = [L]_{\mu} \equiv 
\frac{\p^2 L}{\p v^{\mu} \p v^{\nu}} \dot{v}^{\nu} +
\frac{\p^2 L}{\p v^{\mu} \p q^{\nu}} v^{\nu} -
\frac{\p L}{\p q^{\mu}}\, .
\ee
Helmholtz found as necessary and sufficent conditions on the functions
$G_{\mu}$ of the {\em independent} variables $q, v, \dot{v}$
in order for the Lagrangian to exist \cite{helmholtz}: 
\bea
\label{H1}
\frac{\partial G_{\mu}}{\partial \dot{v}^{\nu}} & = &  
\frac{\partial G_{\nu}}{\partial \dot{v}^{\mu}} \, , \\
\label{H2}
\frac{\partial G_{\mu}}{\partial v^{\nu}} +
\frac{\partial G_{\nu}}{\partial v^{\mu}} & = &
\frac{d}{d t}  \left\{ \frac{\partial G_{\mu}}{\partial \dot{v}^{\nu}} +  
\frac{\partial G_{\nu}}{\partial \dot{v}^{\mu}} \right\} \, , \\
\label{H3}
\frac{\partial G_{\mu}}{\partial q^{\nu}} - 
\frac{\partial G_{\nu}}{\partial q^{\mu}} & = &
\frac{1}{2} 
\frac{d}{d t} \left\{ \frac{\partial G_{\mu}}{\partial v^{\nu}} -  
\frac{\partial G_{\nu}}{\partial v^{\mu}} \right\} \, .
\eea
With respect to an arbitrary time  parameter $t$ 
the  autoparallel equation  (\ref{auto}) is 
\be
\label{autolgs}
G_{\mu} = 
[L_{g}]_{\mu} + 2 S_{\mu\nu\la} \frac{v^{\nu} v^{\la}}{\sqrt{v^2}} = 0 \, .
\ee
It is obvious that the geodesic term 
$[L_{g}]_{\mu}$ fulfills the Helmholtz conditions.
The second term in (\ref{autolgs}) is the torsion force that
causes a deviation of the trajectory from the geodesics
$[L_{g}]_{\mu} = 0$. Due to the linearity in $G_{\mu}$,
the Helmholtz conditions yield restrictions on the torsion force only.
From the second Helmholtz condition (\ref{H2}), 
the restriction $S_{\mu(\nu\la)}=0$ on torsion can be deduced.
This implies vanishing the torsion force in (\ref{autolgs}).
Thus, Eq. (\ref{lg}) does not have any  solution for
a non-vanishing torsion force.

The only possibility to find a Lagrangian formalism
for the autoparallel is to look for an {\em equivalent}
 set of equations which may have the Euler-Lagrange form.
This can be done by introducing a multiplier ${\Omega_{\mu}}^{\nu}(v,q)$ 
with $\det {\Omega_{\mu}}^{\nu} \neq 0$ 
which acts as an integrating factor in Eq. (\ref{lg}).
We are then looking for a solution to the equation
\be
\label{mult}
[L]_{\mu} = {\Omega_{\mu}}^{\nu} G_{\nu} \, .
\ee
The integrability conditions (\ref{H1})-(\ref{H3})
become less restrictive for $G_{\mu}$ itself
since some of them can be fulfilled by an
appropriate choice of the multipliers.
This procedure was first proposed in \cite{havas}.
Although there has been much progress in this approach 
(see \cite{santilli})  and some useful techniques 
have been invented to simplify the Helmholtz conditions,
the problem still remains unsolved.
Recently, the inverse variational problem for Eq. (\ref{auto})
with $F_{\mu}\!=\!0$ has been solved in two dimensions
\cite{anderson}. However, in these works the proper time $s$
in the equation of motion (\ref{auto}) has been considered 
as the Lagrangian time $t$.
Consequently, the actions obtained 
are not time reparametrization invariant and it would
be difficult to give  them a physical interpretation in the
framework of a relativistic theory.
However, they might be useful to study a nonrelativistic 
autoparallel motion on two-dimensional surfaces.

\subsubsection{The gradient case}
\setcounter{equation}0
\label{graca}

Here we show that the Helmholtz integrability conditions can 
be fulfilled for the generalized problem (\ref{mult}).
In the special case when the trace of the torsion tensor is a gradient and the
traceless part vanishes,
\be
\label{grator}
{S_{\mu\nu}}^{\lambda} ={\textstyle \frac{1}{2}} 
(\delta^{\lambda}_{\mu} \; \partial_{\nu} 
\sigma -\delta^{\lambda}_{\nu} \;
\partial_{\mu} \sigma ) \, ,
\ee
the corresponding autoparallel equation (\ref{autolgs}) follows from the 
least action principle $\delta S_{(\s)}$ $= 0$ where \cite{klepel}
\be
\label{actaut}
S_{(\s)} = \int \! L_{(\s)} d t =
 - m \int \!e^{\s(q)} \sqrt{g_{\mu\nu}v^{\mu}v^{\nu}}d t =
-m \int \! e^{\s(q)}  d s \, .     
\ee
Whereas the action (\ref{actgeo}) for geodesics is just an integral
over proper time, in (\ref{actaut})
a scalar factor $e^{\s(q)}$ occurs.
The same Lagrangian was obtained in Brans-Dicke theory
\cite{brans}, where the masses of particles depend on position
$m \rightarrow m(q)=m e^{\s (q)}$. The scalar field $\s$ can also
be interpreted as the dilaton field \cite{fradkin}
emerging in the low energy limit of the string theory together with the
metric $g_{\mu\nu}$.

The Lagrange derivative of $L_{(\s)}$ reads 
\be
\label{ldgt}
[L_{(\s)}]_{\mu} = e^{\sigma} 
\left( [L_{g}]_{\mu} + ( g_{\mu\la} \p_{\nu} \sigma -
g_{\nu\la} \p_{\mu} \sigma ) \frac{v^{\nu} v^{\la}}{\sqrt{v^2}}
\right) = 0 \, .
\ee
Note that the Lagrange derivative has the form
(\ref{mult}) with the  multiplier  
${\Omega_{\mu}}^{\nu}= e^{\sigma} \delta_{\mu}^{\nu}$.
Eq. (\ref{ldgt}) exhibits the time reparametrization symmetry.
The motion can be specified in a gauge invariant way by defining
the proper time.
Since the theory has an extra scalar function $\s$ available,
the gauge invariant time is not unique:
$ds\!=\!f(\s) \sqrt{g_{\mu\nu}v^{\mu}v^{\nu}}\, dt$
with $f(\s)$ being a general positive function of $\s$.
If we set $f=1$, Eq. (\ref{ldgt}) turns into the autoparallel equation
\be
\label{autpt}
g_{\mu\nu} \dot{u}^{\nu} +
\left( \og_{\la\nu\mu} +  
 g_{\mu\la} \p_{\nu} \s -
g_{\nu\la} \p_{\mu} \s  \right) u^{\la} u^{\nu} = 0.
\ee
It should be stressed that the motion depends on the
definition of the (proper) gauge invariant time.
For instance, with the choice $f=e^{\s}$ Eq. (\ref{ldgt})
turns into a {\em geodesic} equation.
Indeed, under the conformal transformation,
\be
\label{conftr}
g_{\mu\nu} \longrightarrow g^{(\s)}_{\mu\nu}= e^{2 \s} g_{\mu\nu} \, ,
\ee
the action (\ref{actaut}) goes over to the action 
(\ref{actgeo}) for geodesics 
associated with the new metric $g^{(\s)}_{\mu\nu}$
and the new proper time $d s^{(\s)}= e^{\s} ds$.
Thus, a violation of Einstein's equivalence principle due to the ``dilaton''
force in (\ref{autpt}) can be observed, provided there is a possibility
to distinguish experimentally between the measurements of distances
and time intervals relative to the metrics $g_{\mu\nu}$ and
$g^{(\s)}_{\mu\nu}$. We return to this issue later in the conclusions.

The metric rescaling (\ref{conftr}) can be used to remove the force caused 
by the ``gradient'' part of the torsion tensor from the equation
of motion:
\be
\label{resem}
G_{\mu} (g_{\alpha\beta},S_{\alpha\beta\gamma}) = 
e^{\sigma} G_{\mu} ( e^{-2 \sigma} g_{\alpha\beta}, 
S_{\alpha\beta\gamma} + S^{(\s)}_{\alpha\beta\gamma}) \, ,
\ee
where $S^{(\s)}_{\alpha\beta\gamma}$ is given by (\ref{grator})
and in both sides of Eq. (\ref{resem}) the proper time is defined
with $f\!=\!1$.

Now we make use of this symmetry 
to built up a quantum field theory which  in a 
semiclassical approximation 
would lead to the autoparallel motion of the
wave packets  in the  ``gradient'' torsion 
and metric background fields.
To this end we go over to the Hamiltonian formalism for the action
(\ref{actaut}). The canonical momenta 
are $p_{\mu}\! =\! \p L_{(\s)} / \p v^{\mu} \!
 = \! - m \, e^{\sigma} v_{\mu} / \sqrt{v^{2}}$, 
so the constraint is
\be
\label{consig}
\Pi_{(\s)} = p^{2} - m^{2} e^{2 \sigma} = 0 \, .
\ee
To construct the corresponding quantum field theory
we can simply adopt the field action (\ref{qftact}) 
with the new metric $g^{(\s)}_{\mu\nu}$ and subject it to quantization.
The correspondence principle is automatically fulfilled.
Indeed, in the semiclassical approximation for the quantum field theory
associated with the action (\ref{qftact})  the wave packets would follow
geodesics with respect to the background metric $g^{(\s)}_{\mu\nu}$
\cite{fierz}.
Making use of the symmetry
(\ref{resem}) we see that the classical trajectories are 
autoparallels with respect to the metric $g_{\mu\nu}$ and the
``gradient'' torsion generated by the background scalar field $\s$.
Thus the scalar field action that leads to a quantum scalar field
theory compatible with the correspondence principle is
\be
\label{qftgra} 
S= \int d^{n}q \; e^{(n-2) \s} \sqrt{g} \left( g^{\mu\nu} \p_{\mu} \phi
 \p_{\nu} \phi - m^{2} e^{2 \s} \phi^{2} \right) \, .
\ee
It yields the following equation of motion for the scalar field $\phi$ 
\be
\label{fieldeq}
\Box \, \phi + (n-2) \, \p_{\mu} \s \, 
\p^{\mu} \phi + m^{2} e^{2 \s} \phi = 0\, .
\ee
where $\Box$ is the Laplace-Beltrami operator: $\Box \phi=(\sqrt{g})^{-1}\,
\p_{\mu} ( \sqrt{g}\; g^{\mu\nu} \p_{\nu} \phi)$.    

Eq. (\ref{fieldeq}) can be regarded as the quantum version of the constraint
(\ref{consig}). Note that a multiplication of the constraint (\ref{consig})
by some function of coordinates would lead to an equivalent constraint on 
the classical level. In quantum theory
the ordering of operators is generally not unique.
Here we have promoted $\Pi_{(\s)}$ into an operator by multiplying it by
$e^{-2 \s}$ and postulating that $e^{- 2 \s} \hat{p}^{2}$
is the Laplace-Beltrami operator with respect to the metric 
$g^{(\s)}_{\mu\nu}$. This ensures the {\em hermiticity}
of the constraint with respect to a scalar product with the measure
$\sqrt{g^{(\s)}} d^{n} q = e^{n \s} \sqrt{g} d^{n} q$,
thus providing the {\em unitarity} of the time evolution.

\subsubsection{Perturbation theory}
\setcounter{equation}0
\label{perturb}

Here we come to the conclusion that there is no 
Lagrangian formalism, except for the gradient case.
We make use of our third physical assumption that the
Lagrangian, if it exists, should be analytical in the torsion tensor.
So far, no experimental observation of torsion has been made.
Therefore the torsion force must be small as compared
with the gravitational force induced by the metric.
This, in turn, suggests solving the integrability conditions
for Eq. (\ref{mult}) by the {\em perturbation} theory in the torsion tensor. 
We shall see that the integrability conditions are not fulfilled even
in first order perturbation theory, thus leading to the
conclusion of the nonexistence of a Lagrangian in general.
We start with the ansatz 
\be 
\label{ansl}
L(v,g,S) = L_{g}(v,g)+ L_{1}(v,g,S) + O(S^{2}) \, ,
\ee
which contains the Lagragian $L_{g}$ (\ref{actgeo}) for geodesics
and a perturbation $L_{1}$ linear in the torsion tensor.
From Eq. (\ref{mult}) follows that the multiplier must also be 
analytic in torsion, so we set
\be
{\Omega_{\mu}}^{\la}(v,g,S) = \delta_{\mu}^{\la}+
{\omega_{\mu}}^{\la}(v,g,S) + O(S^{2})\,  .
\ee
In this approximation, the  substitution of (\ref{autolgs}) in (\ref{mult}) 
leads to
\be
\label{l1lg}
[L_{1}]_{\mu} = 
{\omega_{\mu}}^{\la} [L_{g}]_{\la} +
2 S_{\mu\nu\sigma} \frac{v^{\nu} v^{\sigma}}{\sqrt{v^{2}}} \, .
\ee 
The variables $\dot{v}, v, q$ are considered as independent variables. 
The integrability conditions 
for (\ref{l1lg}) are still difficult to 
analyze because of the presence of the general functions
${\omega_{\mu}}^{\la}$. Therefore we first look for the
integrability conditions in the velocity space assuming the configuration
space point to be fixed.
So we set $q^{\mu}=q^{\mu}_{0}$ after calculating all the derivatives
$\p_{\mu}$ in  (\ref{l1lg}). Eq. (\ref{l1lg}) is covariant
under general coordinate transformations as a consequence of our second
assumption. In particular, we may assume a geodesic coordinate system
\cite{petrov} at $q^{\mu}_{0}$. The advantage of this is that the Christoffel
symbols are zero at the origin ${\og_{\mu\nu}}^{\la}(q_{0})=0$.
Thanks to this property, 
the term ${\omega_{\mu}}^{\la} [L_{g}]_{\la}$ is proportional
to the acceleration $\dot{v}^{\mu}$ and must cancel against the
corresponding term contained in $[L_{1}]_{\mu}$.
This leads to an equation for the multiplier which
is not relevant for the subsequent analysis.
For the remaining terms we obtain
\be
\label{cdtor}
v^{\nu} \frac{\p^{2} L_{1}}{\p q^{\nu} \p v^{\mu}} - 
\frac{\p L_{1}}{\p q^{\mu}} =
2 \,  S_{\mu\nu\sigma} \frac{v^{\nu} v^{\sigma}}{\sqrt{v^{2}}}\, .
\ee
Next, in the vicinity of $q_{0}$ we apply the Fourrier transform 
$L_{1}(q,v)=\int d k \, e^{ikq} \, \tilde{L}_{1}(k, v)$,
similarly for $\tilde{S}_{\mu\nu\s}(k)$, so that 
$\left. \frac{\p L_{1}}{\p q^{\mu}} \right|_{q=q_{0}} =
\int d k \, i \, k_{\mu} e^{i k q_{0}} \tilde{L}_{1}(k,v)$.
Substituting this into (\ref{cdtor}), we obtain a first-order differential    
equation for $\tilde{L}_{1}(k, v)$ as a function of $v^{\mu}$. 
This equation can be simplified by the ansatz
$\tilde{L}_{1}= k_{\mu} v^{\mu}  c(k,v)$, leading to 
\be
\label{glc}
i \frac{\p c}{ \p v^{\mu}} = \frac{2}{(k,v)^{2}} 
 \ts_{\mu\nu\sigma} \frac{v^{\nu} v^{\sigma}}{\sqrt{v^{2}}} \, .
\ee
The integrability conditions for Eq. (\ref{glc})
are now easy to derive.
After multiplying them by the factor 
$\left( (k,v) \sqrt{v^2} \right)^{3}$, they turn into a set of 
vanishing linear combinations of the 
monomials  $v^{\nu} v^{\s} v^{\alpha} v^{\beta}$.
Since $v^{\mu}$  are independent variables we are left with the equation
\be
\label{ks}
2 \, k_{[\mu} \ts_{\la](\nu\sigma} \, \eta_{\alpha\beta)} +
k_{(\nu} \left\{ \ts_{[\mu\la]\sigma} \, \eta_{\alpha\beta)} +
\ts_{\sigma[\la\mu]} \, \eta_{\alpha\beta)} + \ts_{[\la|\alpha\beta}
\, \eta_{\sigma)\mu]} \right\} = 0\, .
\ee
Here the indices $(\nu\s\alpha\beta)$ must be symmetrized, while
the indices in the square brackets $[\mu\la]$ are antisymmetrized.

There are two cases where the integrability condition (\ref{ks})
is identically fulfilled and, hence, the Lagrangian always exists.
First, we observe that $v^{\mu} \p c / \p v^{\mu} \equiv 0$ since
$S_{\mu\nu\s}=-S_{\nu\mu\s}$. Therefore, $c$ depends only on the angular
variables in the velocity space, not on the modulus $\sqrt{v^{2}}$.
In two dimensions, Eq. (\ref{glc}) contains only one non-trivial
equation which always has a solution.
The Lagrangian can be constructed as proposed in Appendix B.
The second case is $\tilde{S}_{\mu\nu\la} \sim \delta^{n}(k)$,
i.e., when the torsion tensor is constant
in the coordinate system chosen. It is easy to obtain a simple
recursion relation for an explicit form of all orders of perturbation theory for the Lagrangian $L$. However, the condition $\p_{\mu} S_{\nu\la\s} =0$
is not covariant under general coordinate transformations.
So, the corresponding Lagrangian is not a scalar and can not be regarded 
as physically acceptable.

The torsion tensor can always be decomposed into a trace,
a totally antisymmetric part and a traceless part $Q_{\mu\nu\la}$ 
which is not totally antisymmetric.
The totally antisymmetric part satisfies (\ref{ks}) identically
because it does not contribute to the torsion force at all.
So we set
\be
\label{tordeco}
{S_{\mu\nu}}^{\s}= {\textstyle \frac{1}{n-1}} 
\left( S_{\mu} \delta_{\nu}^{\s} - S_{\nu} \delta_{\mu}^{\s} \right) +
{Q_{\mu\nu}}^{\s} \, ,
\ee
where $S_{\mu} = {S_{\mu\la}}^{\la}$.
Contracting (\ref{ks}) with  $k^{\nu}k^{\sigma}k^{\alpha}k^{\beta}$,
$\eta^{\nu\sigma} \eta^{\alpha\beta}$ and
$k^{\alpha} k^{\beta} \eta^{\nu\sigma}$ we get a system
\be
\label{concon}
\beta_{\mu\la} + 3 \gamma_{\mu\la}=0 \, , \hspace{1em}
(2n\!+\!5) \alpha_{\mu\la} + (n\!+\!1) \beta_{\mu\la}=0 \, , \hspace{1em}
3 \alpha_{\mu\la} + (n\!+\!4) \beta_{\mu\la} + (2n\!+\!11) 
\gamma_{\mu\la}=0 \, ,
\ee
where $\alpha_{\mu\la} = k_{[\mu} \ts_{\la]}$, 
$\beta_{\mu\la}=k_{\sigma} ( \ts_{[\mu\la]}\,\!^{\sigma} +
\ts^{\sigma}\,\!_{[\la\mu]})$ and
$\gamma_{\mu\la}= k_{[\mu} \ts_{\la]\nu\s} \, 
\frac{k^{\nu} k^{\s}}{k^2}$. 
The determinant of the coefficients is $2(n\!-\!2)(n\!+\!1)$.
So, for $n>2$ we conclude $\alpha_{\mu\nu}=0$ and $\beta_{\mu\nu}=
\gamma_{\mu\nu}=0$.
The first relation gives rise to a restriction on the trace $S_{\mu}$
\be
\label{ks0}
k_{[\mu} \ts_{\la]} =0 \, ,
\hspace{1em}\mbox{hence,}\hspace{1em}
S_{\la}(q) \sim \p_{\la} \s(q) \, .
\ee
That is, in any dimension greater than two the contracted torsion tensor
must be a gradient. We conclude that for a {\em generic} torsion 
the inverse variational
problem for the autoparallel equation has no solution.
The ``gradient'' part of the torsion tensor (\ref{tordeco}) satisfies
(\ref{ks}) identically, so that the integrability condition (\ref{ks})
applies to $Q_{\mu\nu\s}$ only.
We investigate it in three and four dimensions.
Both cases are treated simultaneously.

The tensor $Q_{\mu\nu\s}$ can be parametrized in 3 and 4 dimensions 
respectively as
\be
Q^{(3)}_{\mu\nu\la} = \epsilon_{\mu\nu\sigma} {A^{\sigma}}_{\la} \, ,
\hspace{1em}\mbox{}\hspace{1em}
Q^{(4)}_{\mu\nu\la} = \epsilon_{\mu\nu\sigma\rho} {B^{\sigma\rho}}_{\la} \,  ,
\ee
where $A_{\mu\nu}$ is a symmetric, traceless $3\times 3$ matrix
(since $Q^{(3) \, \nu}_{\mu\nu} = 0$ and $\e^{\mu\nu\s} Q^{(3)}_{\mu\nu\s}=0$),
and $B_{\s\rho\la}$ satisfies
$\e^{\mu\s\rho\la} B_{\s\rho\la}=0$ and must be traceless ${B^{\s\la}}_{\la}=0$
(since $Q^{(4)\, \nu}_{\mu\nu} =0$ and $\e^{\mu\nu\la\s} 
Q^{(4)}_{\nu\la\s}=0$).
Thus, $A_{\mu\nu}$ contains 5 independent components, while 
$B_{\mu\nu\s}$ has 16. They are subject to the conditions 
\bea
\label{cond1}
k_{\sigma} \tilde{A}^{\sigma\rho} = 0 \, ,
\hspace{1em}&\mbox{}&\hspace{1em}
k_{\sigma} \tilde{B}^{[\mu\la]\sigma} = 0 \, ,\\
\label{cond2}
k_{(\nu} \tilde{A}_{\alpha\beta} \delta_{\sigma)}^{\tau} = 0 \, ,
\hspace{1em}&\mbox{}&\hspace{1em}
2 \, \eta_{(\alpha\beta|} k_{\rho} {{\tilde B}^{\rho\!\!\ [ \mu}}
\!\!\ _{|\sigma} 
\delta^{\la]}_{\nu)} + 
k_{(\nu} {\tilde B}_{\beta}\!\!\ ^{[\mu}\!\!\ _{\sigma}
\delta_{\alpha)}^{\la]} = 0 \, . 
\eea
Eq. (\ref{cond1}) is equivalent to $\beta_{\mu\nu}=0$, while Eq. (\ref{cond2})
stems from the integrability condition (\ref{ks}) where $\beta_{\mu\nu}=0$ 
has been taken into account. 
It is possible to select 5 linearly independent equations for $A_{\mu\nu}$
and 16 for $B_{\mu\nu\s}$ out of these equations.
So we conclude that $Q^{(3,4)}_{\mu\nu\la}=0$, and the Lagrangian exists 
only for the ``gradient'' case.

\subsubsection{Conclusions}
\setcounter{equation}0

We have proposed a rather general approach to study possible deviations from
Einstein's equivalence principle due to the coupling between matter and
the torsion of spacetime.
Our approach is based on the inverse problem of the calculus of variations
and general principles of quantum field theory. It is far more general
than the minimal gauge coupling principle which is typically used 
to construct a coupling between matter and spacetime geometry
\cite{kibble,hehl95}.
We have shown that for a generic torsion force which makes the trajectories 
of classical spinless particles the autoparallels, no local quantum 
field theory exists in four dimensions that leads to the autoparallel
motion in the semiclassical (eikonal) approximation. Only when the 
torsion tensor has a special form the above problem admits a solution.
In this case the coupling between matter and torsion is 
equivalently described by the 
dilaton field whose existence is predicted by the string theory \cite{fradkin}.

The Einstein equivalence principle is not violated by the coupling between 
matter and the dilaton field if the coupling obeys the 
{\em universality} principle \cite{polyakov}, meaning that it is constructed by
the replacement $g_{\mu\nu} \rightarrow g^{(\s)}_{\mu\nu} = e^{2 \s} 
g_{\mu\nu}$ in the matter Lagrangian.
Indeed, there would be no experiment that distinguishes the motion of test
particles in the composite background metric $g^{(\s)}_{\mu\nu}$ from
that in the background metric $g_{\mu\nu}$ and the dilaton field $\s$.
A deviation from the Einstein general relativity can only be seen in cosmology
which is affected by {\em dynamics} of the dilaton field 
\cite{polyakov,damour}. 

We remark that the minimal gauge coupling principle does not predict the
dilaton and leads only to the coupling between spin and torsion.
As has been stressed by some authors (see the discussion in \cite{ohanian})
such a coupling might pose a consistency problem since the spin of composed
particles is not simply a sum of the spins
of its constituents, but involves also the orbital
angular momentum. For instance, a spinning particle could be a bound state
of {\em spinless} particles (e.g., a vector boson composed of a few scalar
bosons, etc.). Therefore, such a spinning particle would not interact 
with torsion at all. Thus, when appling the minimal gauge coupling
principle, one has always to decide whether a given kind of particles
is truly elementary or composite. Such a drawback could be circumvented
either by allowing for the coupling of torsion to the angular momentum
or by simply postulating that any theory for composite spinning particles 
should be consistent with the minimal gauge coupling principle,
thus making a restriction on the future fundamental theories.
Given the difficulties of describing composite relativistic
quantum fields, this latter option does not seem easy to pursue
in practice, as well as it does not admit a simple geometrical
interpretation. 

Here we have explored
the first possibility. The autoparallel equation (\ref{auto})
($F_{\mu} \! \equiv \! 0$) can be rewritten as the matter energy-momentum
conservation law \cite{kleinert2}
\be
\label{EMC}
\overline{D}_{\mu} \overline{T}^{\mu\nu} + 2 {S^{\nu}}_{\mu\s} 
\overline{T}^{\mu\s} = 0 \, ,
\ee
where  $\overline{T}_{\mu\nu}$ is the energy-momentum tensor in 
general relativity (see also Appendix A).
The second term in (\ref{EMC}) contains an interaction between the 
torsion and the angular momentum.
We have proved that there exists no local quantum field theory whose dynamics
complies with Eq. (\ref{EMC}) except the special case when all the effects 
of torsion can be interpreted  as those caused by the dilaton.
One could also regard this result as an argument supporting the point 
of view that the spacetime geometry is specified only by the metric and 
possibly by the dilaton field, i.e., by the low energy modes of string theory.

We remark that the minimal price of incorporating Eq. (\ref{EMC}) into
quantum theory is to give up locality \cite{shabanov}. This does
not seem to us acceptable in quantum field theory of fundamental 
interactions, but still may be possible in effective theories describing
a quantum motion of interstitial particles in
crystalls with topological defects.

\vskip 0.2cm
\noindent
{\bf Acknowledgements}

\vskip 0.2cm
\noindent
We both      
acknowledge stimulating discussions with H. Kleinert.
S.V.S. is grateful to F.W. Hehl for useful discussions.
S.V.S. was supported by DFG under the project Kl-25626-1.

\vskip 0.3cm

\noindent
{\bf Appendix A: Energy-momentum conservation for the autoparallels}
\renewcommand{\theequation}{A.\arabic{equation}}
\setcounter{equation}0

\vskip 0.3cm
\noindent
The energy-momentum conservation law follows from the invariance
of the action under general coordinate transformations.
As compared with the geodesic action (\ref{actgeo}), the action (\ref{actaut})
contains an extra scalar field describing the background spacetime geometry
so that its variation is determined by both the variations of the metric 
$g_{\mu\nu}$ and the dilaton $\s$. Thus, we get
\be
\label{ds0}
0=\delta S = \int d^{4} q \, \sqrt{g} \left\{ \frac{1}{2} T^{\mu\nu}
\delta g_{\mu\nu} + {T^{\mu}}_{\mu} \, \delta \s
+ \frac{\delta {\cal L}}{\delta q^{\mu}} \, \delta q^{\mu} \right\} \, ,
\ee
where, as usual,
\be
\label{emt}
T^{\mu\nu}(q) \equiv \frac{\delta {\cal L}}{\delta g_{\mu\nu}(q)}
= \frac{1}{\sqrt{g(q)}} \int \! d s \, \delta^{4} \, (q-q(s))
\,  p^{\mu} u^{\nu}  
\ee
is the energy-momentum tensor and
${\cal L}$ is the Lagrangian density defined by $L= \int d^{3}q \, {\cal L}$. 
We observe that $T^{\mu\nu} = e^{\s} \overline{T}^{\mu\nu}$
where $\overline{T}^{\mu\nu}$ is the energy-momentum tensor
for the geodesic motion.
The difference occurs through the $\s$-dependence of the particle momentum:
$p^{\mu}=-m e^{\s(q)} u^{\mu}$.
One can easily convince oneself that $\delta {\cal L} / \delta \s = 
{T^{\mu}}_{\mu} \equiv T$, which specifies the second term in (\ref{ds0}).
For the actual motion of the particle the third term in
(\ref{ds0}) vanishes and we get the energy-momentum conservation law
(cf. (\ref{EMC}))
\be
\overline{D}_{\mu} \overline{T}^{\mu\la} +
\overline{T}^{\mu\la} \p_{\mu} \s 
- \overline{T} \p^{\la} \s  = 0 \, .
\ee
So we see two additional terms occurring in the conservation law 
due to the torsion force. 
Integrating this equation over a three-dimensional spacelike hypersurface
$q^{0}=const$ we again recover the autoparallel equation for the
``gradient'' torsion (\ref{autpt}).

\vskip 0.3cm
\noindent
{\bf Appendix B: The autoparallel Lagragian in two dimensions}
\renewcommand{\theequation}{B.\arabic{equation}}
\setcounter{equation}0

\vskip 0.3cm
\noindent
In two dimensions the integrability conditions (\ref{ks}) yield no
restriction on torsion because of the time reparametrization invariance.
Indeed, by fixing the gauge $q^{0} \equiv t$ the problem becomes
one-dimensional. A general solution of the one-dimensional
inverse variational problem was found by Darboux \cite{darboux}.
So, the Lagrangian always exists for the two-dimensional 
autoparallel motion.
However, the constraint appears to be non-polynomial in the canonical
momenta, thus leading to a {\em nonlocal} quantum field theory.

The torsion tensor can be parametrized in two dimensions  
by two scalar functions $\la$ and $\s$:  
\be
\label{tor2d}
{S_{\mu\nu}}^{\alpha} = 
{\textstyle \frac{1}{2}} \,  \epsilon_{\mu\nu} \left( \p^{\alpha} \la + 
\epsilon^{\alpha\beta} \p_{\beta} \sigma \right).
\ee
Setting $\varphi=(k,u)/ \sqrt{k^{2}}$,
we may decompose  $u^{\mu} = [ \varphi \; k^{\mu} + 
( 1-\varphi^{2} )^{\frac{1}{2}}
 \; (\epsilon k)^{\mu} ] / \sqrt{k^{2}}$.
Solving Eq. (\ref{glc}) for $c=c(\varphi)$ we find
\be
i \tilde{L}_{1} = \sqrt{v^{2}}  \; \left(  \tilde{\s} + 
\varphi \ln [ \varphi^{-1}  + ( \varphi^{-2}-1 )^{\frac{1}{2}}  ]
\; \tilde{\la} \right) \, .
\ee  
The first term is the linear part of the Lagrangian (\ref{actaut})
for the ``gradient'' torsion.
The second term is  non-polynomial in
$\varphi$. 
This fact is not changed by the higher orders of the  
expansion  (\ref{ansl}) as can be seen from a recursion relation 
for $L_{i}$.
Because of this, the Lagrangian would lead to a constraint which is  
non-polynomial in $p$. Thus the corresponding quantum field theory
would be non-local. So, we conclude that also 
in two dimensions only the ``gradient'' torsion
leads to an acceptable theory.

It is certainly possible to find a Lagrangian for the generic
torsion (\ref{tor2d}). However a complete discussion would be too 
involved and goes beyond the scope of this letter.
Just to give an idea of how the Lagrangian would look like,
we calculate it under the simplifying conditions 
$g_{\mu\nu}=\eta_{\mu\nu}$ and $\p_{0} \la =0$.
This latter condition obviously violates the general coordinate invariance,
but will allow us to find an {\em explicit} form of the Lagrangian.
We also set $\s=0$ since the gradient case has been already discussed.
After fixing the gauge by $q^{0}=t$  ($v^{0}=1$)
and adopting the notations   $v^{1} \equiv v$, 
$\p_{1} \la \equiv \p_{x} \la$
we get one simple equation out of the autoparallel equation (\ref{auto})
\be
\label{aut1d}
\dot{v} + \p_{x} \la \, ( v- v^{3} ) =0 \, .
\ee
The associated Lagrangian can be found via 
the Hamiltonian formalism.
The first Hamiltonian equation is set to be $\dot{x} = p / \sqrt{1+p^{2}} 
= \omega \p_{p} H$. Then the second Hamiltonian equation can be derived
from (\ref{aut1d}) as $\dot{p} =\! - p \, \p_{x} \la =  - \omega \p_{x} H$.
These are equations for the Hamiltonian $H$ and the symplectic structure
$\omega$ which can easily be solved. Next,
choosing Darboux coordinates $X=x$ and 
$P$ such as $\p_{p} P  = \omega ^{-1}$, the Lagrangian is
obtained by the Legendre transformation for $P$:
$L(v,\la) = P \dot{x} -H$.
The time reparametrisation invariance is restored 
by the rule $L_{(\la)}(v^{0}, v^{1}, \la) = 
v^{0} L(v^{1} / v^{0}, \la)$  \cite{diraclec}.
Therefore the Lagrangian is:
\be
\label{l2d}
L_{(\la)} = 
- \sqrt{v^{2}} \cosh{\la}
- v^{1} \, \ln C \;
\sinh{\la} \, ,
\ee
where $C=| \, v^{0}  / v^{1} + \sqrt{   
( v^{0} / v^{1} )^{2} - 1} \, |$.
It is not difficult to see that the constraint resulting from $u^{2}\!=\!1$
is not polynomial in the canonical momenta $p_{\mu}$ because 
$p_{\mu}\!=\!p_{\mu}(u)$ contain $\ln C$.

\end{document}